\documentclass[twocolumn]{aastex6}
\newcommand{\pdif}[2]{\frac{\partial #1}{\partial #2}}

\shortauthors{Wakita et al.}

\begin{document}
\title{Diffusion of oxygen isotopes in thermally evolving planetesimals and size ranges of presolar silicate grains}

\author{Shigeru Wakita\altaffilmark{1}, Takaya Nozawa\altaffilmark{2}, and Yasuhiro Hasegawa\altaffilmark{3}}

\altaffiltext{1}{Center for Computational Astrophysics, National Astronomical Observatory of Japan, Mitaka, Tokyo 181-8588, Japan; shigeru@cfca.jp}
\altaffiltext{2}{Division of Theoretical Astronomy, National Astronomical Observatory of Japan, Mitaka, Tokyo 181-8588, Japan}
\altaffiltext{3}{Jet Propulsion Laboratory, California Institute of Technology, Pasadena, CA 91109, USA}

\begin{abstract}
Presolar grains are small particles found in meteorites through their isotopic compositions which are considerably different from those of materials in the Solar System. 
If some isotopes in presolar grains diffused out beyond their grain sizes when they were embedded in parent bodies of meteorites, 
their isotopic compositions could be washed out, and hence the grains cannot be identified as presolar grains any more. 
We explore this possibility for the first time by self-consistently simulating the thermal evolution of planetesimals and the diffusion length of $^{18}$O in presolar silicate grains.
Our results show that presolar silicate grains smaller than $\sim 0.03 \micron$ cannot keep their original isotopic compositions even if the host planetesimals experienced maximum temperature as low as 600 $^{\circ}$C. 
Since this temperature corresponds to the one experienced by petrologic type 3 chondrites, the isotopic diffusion can constrain the size of presolar silicate grains discovered in such chondrites to be larger than $\sim 0.03 \micron$.
We also find that the diffusion lengths of $^{18}$O reach $\sim 0.3-2 \micron$ in planetesimals that were heated up to 700-800 $^{\circ}$C. 
This indicates that, if the original size of presolar grains spans a range from $\sim 0.001 \micron$ to $\sim 0.3 \micron$ like that in the interstellar medium, the isotopic records of the presolar grains may be almost completely lost in such highly thermalized parent bodies. 
We propose that isotopic diffusion could be a key process to control the size distribution and abundance of presolar grains in some types of chondrites.
\end{abstract}

\keywords{meteorites, meteors, meteoroids -- planets and satellites: formation -- diffusion}

\section{Introduction} \label{intro}
Primitive meteorites contain unique tiny materials called presolar grains. The fundamental property of the presolar grains is their isotopic compositions, which are largely deviated from extremely homogeneous values of materials in the Solar System.
In particular, presolar silicate grains are identified via the oxygen isotopic ratios of $^{17}$O/$^{16}$O and $^{18}$O/$^{16}$O \citep[e.g.,][]{cn04}.
Since such isotopic anomalies have to be preserved over the entire history of the Solar System (that's why we can currently measure the differences in isotopic composition), investigation of presolar grains can 
provide us with important clues to understand the formation and evolution of the Solar System.

It is considered that the presolar grains originally formed in nearby stars at the post-main sequence phases such as supernovae and/or asymptotic giant branch (AGB) stars. 
Then they were transported to the presolar nebular that is a pre-stage of forming the Sun,
and were finally incorporated into planetesimals that are parent bodies of meteorites.
Some of theoretical studies suggest that dying stars could inject
relatively large grains with radii of 0.1-1 $\micron$ into the interstellar medium \citep[ISM, e.g.,][]{nkh07,yk12}.
On the other hand, while they were traveling to the presolar nebula, many of them might be fragmented into grains smaller than $0.1 \micron$ due to shattering in the interstellar turbulences \citep{hy09}.
It is nonetheless important to point out that such small ($<0.1 \micron$) presolar grains have been rarely detected in the currently available samples; the typical size distribution of presolar silicate grains spans the range from $\sim$ 0.1 to $\sim$ 1 $\micron$ \citep[e.g.,][]{z03,hg09,nns10,lvh12,hlk15}.

The abundance of presolar grains varies among different petrologic types of meteorites: 
in general, presolar grains are the most abundant in type 3, and as the type number increases from 3 to 6, the abundance of presolar grains decreases \citep{h90,hl95}.
These types are based on the degree of metamorphism experienced 
by meteorites when they were embedded in planetesimals. 
Since the metamorphism depends on the temperature, petrologic types are regarded as representing a peak temperature experienced by the planetesimals.
For instance, it is widely accepted that the petrologic type 3 ordinary chondrites experienced a peak temperature less than 700 $^{\circ}$C, type 6 did higher than 800$^{\circ}$C, and types 4 and 5 did between them \citep{hrg06}.
Thus, this classification suggests that the abundance of presolar grains may be related to the thermal history of planetesimals.

It is implicitly presumed that the metamorphism totally erases the isotopic records of presolar grains. 
This may be because the metamorphism creates new minerals or crystalline structures by breaking the original atomic bonds and forming new bonds. 
If the metamorphism would be the dominant process to wash out the isotopic compositions, more presolar grains would be discovered in unmetamorphosed (primitive) chondrites.
While the abundance of presolar grains in the least metamorphosed type 3 chondrites have the highest abundance of presolar grains among chondrites \citep{nsz07,fs09,nns10,fs12,nas13}, it is still much lower than that in interplanetary dust particles (IDPs) regarded as the most primitive materials \citep{mksw03,fsb06,bnc09}.
Therefore, the metamorphism would not explain the difference in abundance of presolar grains between type 3 chondrites and IDPs.
On the other hand, even if chemical compositions of minerals are not changed through the metamorphism, the replacement of an atom with another one, so-called atomic diffusion, could be realized in slightly thermalized planetesimals. 
Therefore, it can be anticipated that the atomic diffusion can also delete the original isotopic composition possessed by presolar grains in their parent bodies.

In this paper we explore this possibility for the first time by computing the diffusion length of $^{18}$O in presolar silicate grains in meteorites and by comparing it with the actually measured size of presolar grains.
Since the diffusion length is sensitive to the temperature, we numerically simulate thermal evolution of  planetesimals with different radii and formation times. 
We find that the diffusion process of oxygen atoms can regulate the size distribution of presolar silicate grains and that only the grains larger than $\sim 0.3 \micron$ can keep their original isotopic properties in types 4-6 chondrites.
Our results also suggest that, even in type 3 chondrites that experienced only a peak temperature of 600 $^{\circ}$C, 
the atomic diffusion can entirely erode the isotope records of presolar silicate grains smaller than $\sim 0.03 \micron$.
This may be viewed as a potential explanation of why the currently available samples of presolar silicate grains in type 3 chondrites have a  size larger than $\sim 0.05 \micron$.
Thus, we conclude that isotopic diffusion is one of the important processes to govern the survival of presolar grains in thermally evolving planetesimals.

\section{Isotopic diffusion in thermally evolving planetesimals} \label{mod}

In order to compute the diffusion length of oxygen isotopes in presolar silicate grains, we adopt a diffusion coefficient of $^{18}$O in olivine from \citet{dcb02}.
The diffusion coefficient $D(T)$ is given as
\begin{equation}
  D(T) = 10^{-8.34}\exp(-3.38\times10^{5}/RT), \label{eq:dc}
\end{equation}
where $R$ is the gas constant [J/K/mol] 
and $T$ is the temperature [K].
Since the diffusion coefficient is obtained for a temperature range of 1100 $^{\circ}$C $<T<$ 1500 $^{\circ}$C \citep{dcb02}, we extrapolate the results down to lower temperatures.
Measurements with the transmission electron microscope show that the chemical compositions of presolar silicate grains vary on the scale of a few tens nanometer \citep{nsz07,bnc09,lvh12}.
While it would be interesting to see how the diffusion coefficients depend on stoichiometry of elements in silicate, there is no data for silicate compositions with a variety of Mg/O, Fe/O and Si/O ratios.
Thus, we assume that grains have homogeneous composition of Mg-rich forsterite used in \citet{dcb02}.

As can be seen from Equation (\ref{eq:dc}), the diffusion coefficient strongly depends on the temperature. 
Thus, we need to numerically simulate thermal evolution of planetesimals to reliably estimate the diffusion length of $^{18}$O in silicate. 
We assume that decay energy of short-lived radioisotope $^{26}$Al heats up materials within a planetesimal as most thermal modelling studies did \citep[e.g.,][]{mft82}. 
A heat conduction equation, 
\begin{equation}
  \rho c \pdif{T}{t} = \frac{1}{r^2}\pdif{}{r} \left( r^2 K \pdif{T}{r} \right) + A \exp(-\lambda t) \label{eq:heat},
\end{equation}
is solved numerically based on \citet{wni14},
where $t$ is time measured from the formation time of planetesimals $t_0$, $r$ is the distance from the center of the planetesimals, $A$ is the radiogenic heat generation rate per unit volume, and $\lambda$ is the decay constant of the radionuclides.
We adopt physical parameters of thermal conductivity $K$= 2 [J/s/m/K], density $\rho$= 3300 [kg/m$^3$] and specific heat $c$ = 910 [J/kg/K] \citep{ym83,ocb10}.
We assume that these parameters do not depend on the temperature.
Radius and formation time of planetesimals, both of which are expected to affect a maximum temperature achieved by planetesimals \citep[see Figure 8 in][]{wni14}, are parameterized; 
the formation time of planetesimals $t_0$ is determined based on a formation time of Ca-Al-rich inclusions (CAIs), 
 4567 Myr ago, when they had the initial ratio of $^{26}$Al/$^{27}$Al $=5.25 \times 10^{-5}$ in the solar nebula \citep[e.g.,][]{cbk12}. 
The radiogenic heat generation rate $A$ falls in direct proportion to the initial ratio of $^{26}$Al/$^{27}$Al, which depends on the formation timing of planetesimals $t_0$ \citep[see Figure 15 in][]{ws11}; 
since the abundance of $^{26}$Al decreases with time (half-life of $^{26}$Al is 0.72 Myr), a planetesimal formed at a later time has less heating sources than that formed earlier.

A diffusion length ($L$) is calculated as $dL^2 = D(T)dt$ at each time step during the thermal evolution of planetesimals.
It takes about $10^3$ years to diffuse $^{18}$O entirely in particles with the size of 1 $\micron$ at 1000 $^{\circ}$C and a much longer time ($10^9$ years) is needed at 600 $^{\circ}$C.
These times are long enough compared with the time step for calculating the thermal evolution of planetesimals ($dt$), which is an order of one year.
We compute the cumulated diffusion length of $^{18}$O given by $L^2 = \sum_{t} dL^2$.
We assume that planetesimals do not experience any kind of disruptions after the formation.

\begin{figure*}[t]
\figurenum{1}
\plotone{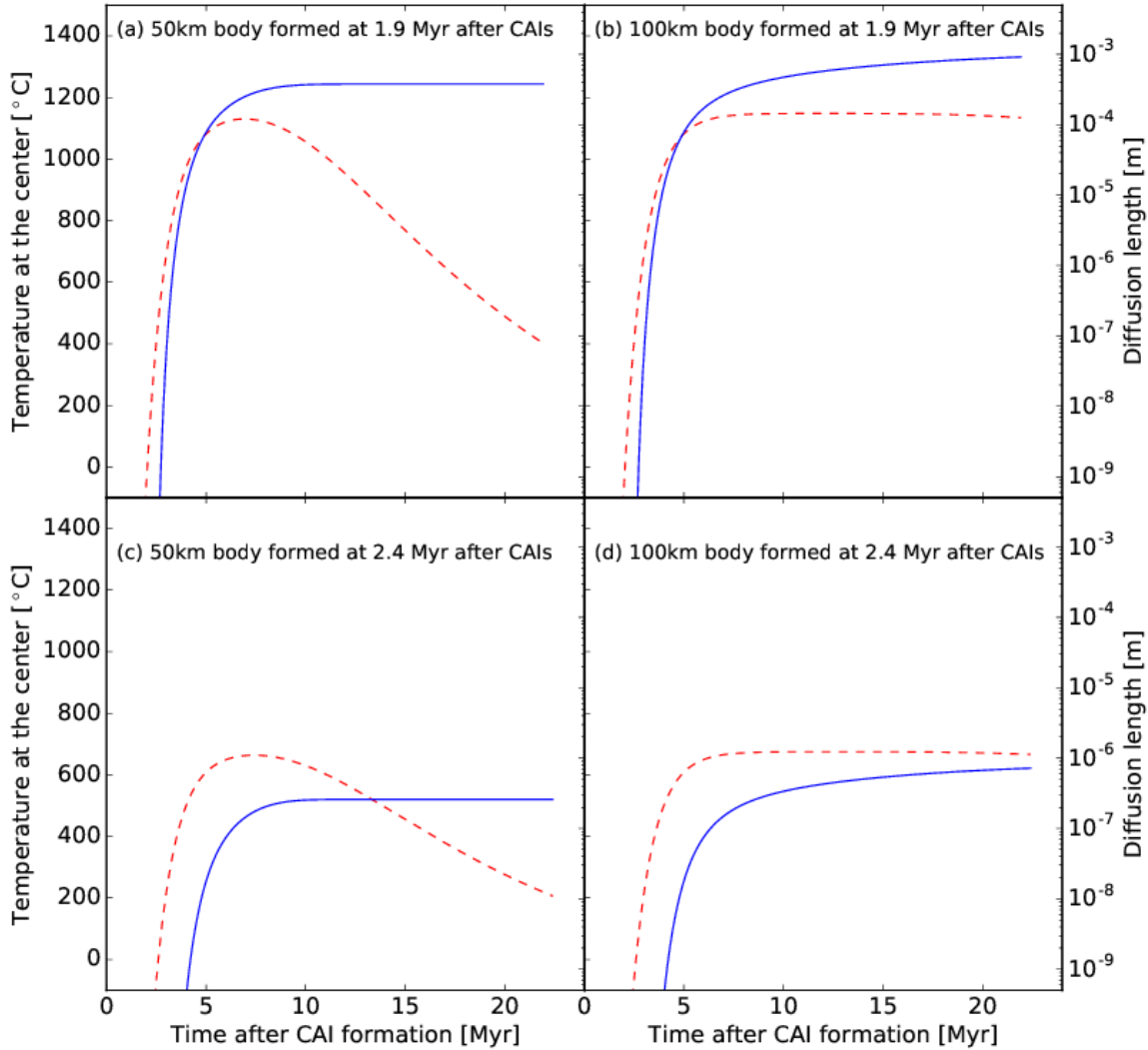}
\caption{Temperature evolutions at the center of planetesimals (dashed lines with left axes)  and the corresponding evolutions of diffusion lengths of $^{18}$O in olivine (solid lines with right axes) as a function of time after CAI formation. 
Each panel gives the result for planetesimals with radii of (a) 50 km and (b) 100 km formed at 1.9 Myr after CAI formation, and with radii of (c) 50 km and (d) 100 km formed at 2.4 Myr after CAI formation. \label{fig1}}
\end{figure*}

Figure \ref{fig1} shows the temperature evolutions at the center of planetesimals (dashed lines) and the accumulated diffusion length of $^{18}$O (solid lines) for planetesimals with different radii (50 km and 100 km) and formation times ($t_0=$ 1.9 Myr and 2.4 Myr). 
The results show that the temperature evolution controls the time development of diffusion length: as the temperature increases, the diffusion length becomes longer. 
When the temperature starts to decrease, the increase in diffusion length ceases (see the left panels of Figure. \ref{fig1}).
Since we evaluate the cumulated diffusion length, it does not decrease after temperature starts to drop. 
This means that the maximum temperature governs diffusion length of $^{18}$O in grains. 

Our results also indicate that for larger planetesimals (see the right panels of Figure \ref{fig1}), the diffusion length at the center gradually increases even after the temperature reaches the maximum value. 
This arises because larger planetesimals can keep the maximum temperature for a longer time than smaller ones (see Figure \ref{fig1} (b) and (d)). 
It should be noted that the maximum temperature, accordingly diffusion length, is more sensitive to the formation times of planetesimals than their radii. 
As seen from Figure \ref{fig1}, the diffusion lengths of planetesimals formed at earlier times ($t_0$ = 1.9 Myr, top panels of Figure \ref{fig1}) are much longer ($> \sim10^{-4}$ m) than those ($< \sim 10^{-6}$ m) formed at later times ($t_0$ = 2.4 Myr, bottom panels of Figure \ref{fig1}). 
This is because, if their sizes are the same, the planetesimals formed at earlier epochs have more abundant $^{26}$Al and reach a higher maximum temperature.
Note that the above results are obtained at the central region of planetesimals. The diffusion lengths are different at different radii of planetesimals.
As mentioned above, the resultant diffusion length is a function of only the maximum temperature. 
Therefore, we can estimate a diffusion length at every location of planetesimals by simply referring to the maximum temperature achieved there.

\begin{figure}[t]
\figurenum{2}
\plotone{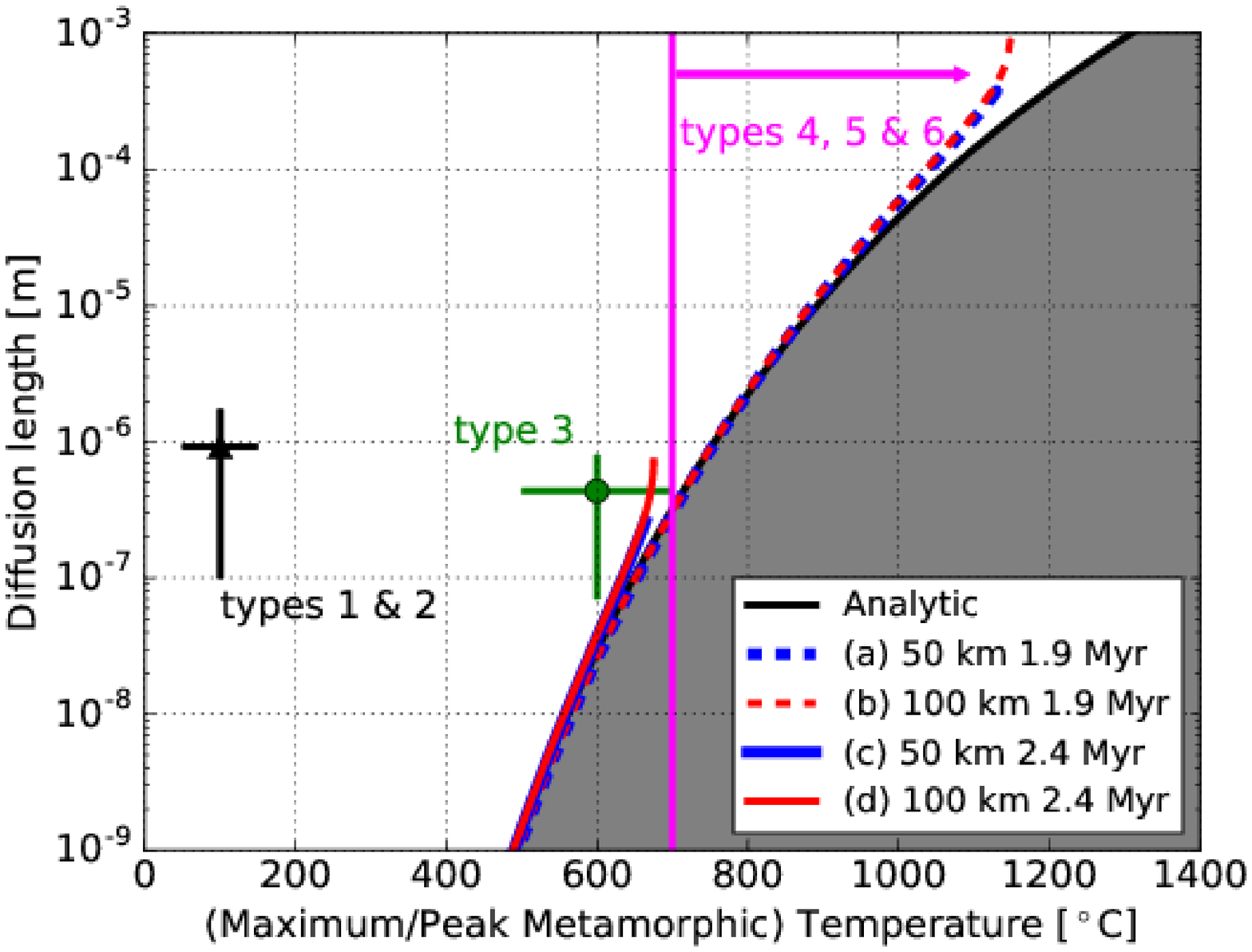}
\caption{Diffusion length of $^{18}$O as a function of maximum temperature, which can be translated to radial depth of the planetesimal. Numerical results are denoted by colored lines, while the analytical one by the black one. For the radii and formation times of planetesimals, we adopt the same values as in Figure \ref{fig1} and the parameter set for each line is given in the legend of the figure. 
The shaded region denotes a regime where diffusion can lead to complete erosion of presolar grains with a certain size. 
Symbols with bars plot the ranges of size and peak temperature derived for presolar silicate grains in type 3 (green-cricle) and types 1-2 chondrites (black-triangle).
The vertical magenta line with a rightward arrow indicates the lower value of the peak temperatures expected for types 4-6 chondrites.
 \label{fig2}}
\end{figure}

We also find that the diffusion length of  $^{18}$O can be approximately described as a following analytical formula,
\begin{equation}
  L^2 = D(T_{\rm max}) {\Delta t_{\rm max}}, \label{eq:analitic}
\end{equation}
where $\Delta t_{\rm max}$ is the duration time of the maximum temperature $T_{\rm max}$. 
Figure \ref{fig2}  represents the diffusion lengths obtained from numerical simulations and from Equation (\ref{eq:analitic}) with $\Delta t_{\rm max}=1$ Myr as a function of maximum temperature. 
It should be emphasized that the dependence of the diffusion length on the maximum temperature follows a simple relation and is reasonably described as a function of $T_{\rm max}$ by Equation (\ref{eq:analitic}).
The maximum temperature achieved at each location of planetesimals decreases with increasing radius.
Thus, each curve in Figure \ref{fig2} can be viewed as representing the radial dependence of diffusion length, where the longest diffusion length is achieved at the center of planetesimals.
While we adopt $\Delta t_{\rm max} = 1$ Myr in Equation (\ref{eq:analitic}), which is a typical duration time of planetesimals with 50 km in radius, it works well for planetesimals with the radius of 100 km. 
Thus, our results demonstrate that 
our analytical formula (Equation (\ref{eq:analitic})) reproduces the numerical results very well and the maximum temperature can be used as an indicator to estimate the diffusion length of $^{18}$O.

\section{Discussion and conclusions} \label{disc}

Here we discuss the implications of our results for the size distribution and abundances of presolar grains in various petrologic types of meteorites. 
In Figure \ref{fig2}, we compare the calculated diffusion lengths with the size ranges of presolar silicate grains obtained from meteorites. 
For two data points on grain sizes, the vertical error bars cover the maximum and the minimum sizes, and the horizontal ones are the suggested ranges of peak metamorphic temperatures.
The sizes of presolar silicate grains are 
0.1-1.7 $\micron$ and 0.07-0.6 $\micron$ in petrologic types 1-2 and 3 chondrites, respectively \citep{z03,hg09,nns10,lvh12,hlk15}.
The metamorphic temperature of type 3 chondrites is likely to be between 500 $^{\circ}$C and 700 $^{\circ}$C \citep{hrg06, kyh07}, whereas that of types 1 and 2 carbonaceous chondrites might be around 100 $^{\circ}$C \citep{kna15}. 

The maximum temperature governs the diffusion lengths.
The length is $\sim 0.001 \micron$ when the maximum temperature is 500 $^{\circ}$C. 
If planetesimals reach the maximum temperature of 700 $^{\circ}$C, the diffusion length is $\sim 0.3 \micron$.
Hence, the diffusion length of $^{18}$O in silicate grains in type 3 chondrites ranges from $\sim 0.001 \micron$ to $\sim 0.3 \micron$, 
depending on the maximum temperature that they experienced.
This indicates that, once we know the minimum size of presolar silicate grains, we can estimate the maximum temperature of their parent bodies.
For example, chondrites which contain the smallest presolar silicate grains of 0.07 $\micron$ would not experience a peak temperature higher than $630 ^{\circ}$C when they were embedded in their parent bodies.
Our calculations imply that, even if typical type 3 chondrites have undergone the peak temperature of $600 ^{\circ}$C, presolar grains smaller than $\sim 0.03 \micron$ can lose their oxygen isotopic anomaly due to the isotopic diffusion. 
This may be one of the reasons why we cannot discover such small presolar silicate grains in type 3 chondrites.

Petrologic type 3 chondrites could be subdivided into types 3.0 to 3.9 based on their characteristics of thermoluminescence and compositions of minerals \citep{sgmr80,sj90,sjr94,gb05,bbm06}.
It is suggested that type 3.0 and 3.1 chondrites experienced lower peak temperatures of $\sim$ 200 - 400 $^{\circ}$C than typical type 3 chondrites \citep{cay08}. 
If this is true, our results imply that the original abundance of presolar silicate grains would be kept in such type 3.0 and 3.1 chondrites. 
In fact, it is interesting to notice that a type 3.0 chondrite, ALHA 77307 is placed as one of primitive chondrites having the highest abundance of presolar silicate grains among type 3 chondrites \citep{nsz07}.
Although the subtypes of type 3 chondrites cannot be clearly distinguished by their peak metamorphic temperatures \citep[e.g.,][]{hrg06}, it would be valuable to see how the abundance of presolar silicates changes in each subtype of type 3 chondrites.

If the birth places of presolar silicate grains are the outflowing gas from supernovae and AGB stars, their initial sizes could be dominantly in a range from 0.1 to 1 $\micron$ \citep[e.g.,][]{h08, nkh07,nwh15}.
As seen in Figure \ref{fig2}, an upper end ($\sim 1 \micron$) for the size of presolar silicate grains roughly corresponds to the maximum size of grains that could form in supernovae and AGB stars.
This implies that the largest presolar grains that might be produced in stellar sources can survive any destructive events \citep[e.g.,][]{hna16}
and they may contain the intact information that was recorded at the time of their formation.
On the other hand, many of the submicron-sized grains collide with each other and fragment into nanometer-sized grains due to shattering in the interstellar turbulence \citep{hnyk10,athn14}. 
Hence, presolar grains should also include grains smaller than $0.1 \micron$.
Nontheless, the size ($\sim 0.1 \micron$) of the smallest silicate grains measured in type 3 chondrites is much larger than the minimum size ($\sim 0.001 \micron$) of interstellar dust.
In fact, the critical value of $0.1 \micron$ coincides with a cumulated diffusion length expected from our results.
This may suggest that the smallest size of presolar grains in type 3 chondrites might be determined by the diffusion process triggered in thermally evolving planetesimals.

For types 1 and 2 chondrites, which experienced only the peak metamorphic temperatures of $\sim 150 ^{\circ}$C or less, the expected diffusion lengths of $^{18}$O are extremely short. 
Therefore, any size of presolar silicate grains 
can survive against the isotopic diffusion.
However, the sizes of presolar grains measured in these petrologic types are confined to a range from 0.1 to 1.7 $\micron$, and grains smaller than 0.1 $\micron$ have not been discovered so far. 
In general, in types 1 and 2 chondrites, the aqueous alternation may have played an important role in eliminating the original isotopic information of grains smaller than $0.1 \micron$.
\citet{lvh12} suggests that an initial abundance of presolar silicate grains might be 10 times larger than current values in a type 2 chondrite and this reduction may be due to destruction by aqueous alteration.

It should be kept in mind that there is a limitation of the size measurement of presolar grains.
Most of the data on the size of presolar grains are obtained from \textit{in situ} measurements of a secondary ion mass spectrometry (SIMS), whose beam size is comparable to a size of $0.15 \micron$.
This indicates that significant amounts of presolar grains with that size or smaller could be undetected.
An optimized setting of NanoSIMS makes it possible to identify smaller ($<  0.1  \micron$) presolar grains \citep{hlk15}.
However, the measurements with spatial resolution of $0.1\micron$ might suffer from instrumental biases in detecting smaller grains \citep{nsz07,nns10}.
Thus, more sophisticated techniques are desired to find
presolar silicate grains smaller than 0.1 $\micron$ in types 1-3 chondrites 
and to check our scenario in which the isotopic diffusion (and aqueous alteration) can affect the size distribution of presolar grains surviving in planetesimals.

In types 4, 5 and 6 chondrites, which are considered to have experienced the peak metamorphic temperature above 700 $^{\circ}$C (the vertical magenta line with a rightward arrow in Figure \ref{fig2}), the expected diffusion length is comparable to or even higher than $\sim 0.3 \micron$.
Therefore, if the original size distribution of presolar grains is limited below $\sim 0.3 \micron$ following the grain size distribution in the ISM \citep{mrn77,nf13}, all of the presolar grains cannot keep their original isotopic compositions
due to the isotopic diffusion.
Note that both the isotopic diffusion and thermal metamorphism may be able to delete the original information of presolar grains in these highly thermalized chondrites.
Nontheless, if the isotopic diffusion is a more efficient process to affect their original isotopic composition than the thermal metamorphism,
then there is a chance to find a large presolar grain of $> 1 \micron$ in types 4-6 chondrites.
As demonstrated above, their minimum sizes have valuable hints to the maximum temperatures experienced by the parent bodies.
Therefore, the search for presolar grains in such fully metamorphosed meteorites is highly encouraged.

Finally, it would be worth discussing the abundances of presolar grains in chondrites and interplanetary dust particles (IDPs). 
The abundances of presolar grains in IDPs are higher than those in chondrites \citep{mksw03,fsb06,bnc09}.
It is very likely that IDPs, which have never experienced any kinds of metamorphisms, can keep the most primitive information about the chemical composition in the solar nebula. 
Thus, it may be plausible to consider that the abundance of presolar grains in IDPs represents the original abundance of presolar grains in meteorites.
There should be some processes to reduce the abundances of presolar grains in chondrites: 
aqueous alteration and thermal metamorphism.
Although these processes can explain the abundances in types 1-2 and 4-6 chondrites, respectively, they cannot be effective for type 3 chondrites.
Hence, the isotopic diffusion may be a primary process to cause the difference in abundance of presolar grains between
primitive chondrites and IDPs.

As we examine the diffusion of $^{18}$O in thermally evolving planetesimals, 
we can also apply the same approach to the diffusion process of other atoms.
There are two interesting presolar silicate grains ($\sim 0.2 \micron$) that were found in an ungrouped carbonaceous chondrite (types 2 or 3) and have a few nanometers of iron-rich rims \citep{fs12}.
This discovery suggests that these rims might be an outcome of a kinetic process or diffusion process.
We estimate the diffusion length of iron using Equation (\ref{eq:analitic}) with the diffusion coefficient of iron in olivine, which is described in terms of temperature, oxygen fugacity and iron content \citep{mma02,dc07}.
At any temperature, the diffusion coefficient of iron would be much larger than that of $^{18}$O.
Thus, iron can easily diffuse into whole grains while oxygen keeps their original content.
When we assume iron content of Fe/(Fe+Mg)=0.5, the diffusion length of iron would be on the order of a nanometer when the maximum temperature is about $200^{\circ}$C.
Hence, the diffusion process could explain a thin Fe-rich rim around the presolar grains with keeping their original oxygen isotopic composition when they were in such a hardly heated parent body.

In this paper, we have examined an isotopic diffusion process in thermally evolving planetesimals, which has never been 
investigated carefully. 
We find that the diffusion can be viewed as an important process to wash out the original isotopic composition of presolar silicate grains in certain meteorites.
We show that the isotopic diffusion can regulate the lower size limits of presolar silicate grains 
in various petrologic types of chondrites, while the upper limits are probably originated from formation processes of the grains in stellar envelopes.
For carbonaceous chondrites (usually types 1-3), measurements of $< 0.1 \micron$ grains are needed to firmly address how
important the diffusion is in reforming the size distribution of presolar grains in planetesimals. 
The methodology developed in this paper is applicable for any other isotopes in variety of minerals
provided that their diffusion coefficients are given.
The measured sizes of presolar grains, combined with the simulations of diffusion lengths of the relevant isotopes, will surely advance the understanding of presolar grains and parent bodies of their host meteorites.

\acknowledgments
We thank an anonymous referee for helpful comments and suggestions.
Numerical computations were carried out on PC cluster at Center for Computational Astrophysics, National Astronomical Observatory of Japan.
S. W. thanks Hiroyuki R. Takahashi for discussions about numerical simulations.
T. N. has been supported in part by a JSPS Grant-in-Aid for Scientific Research (26400223).
The part of this research was carried out at JPL/Caltech under a contract with NASA. Y. H. is supported by JPL/Caltech.


\begin{thebibliography}{}
\bibitem[Asano et al.(2014)]{athn14} Asano, R.~S., Takeuchi, T.~T., Hirashita, H., \& Nozawa, T.\ 2014, \mnras, 440, 134

\bibitem[Bonal et al.(2006)]{bbm06} Bonal, L., Quirico, E., Bourot-Denise, M., \& Montagnac, G.\ 2006, \gca, 70, 1849 
\bibitem[Busemann et al.(2009)]{bnc09} Busemann, H., Nguyen, A.~N., Cody, G.~D., et al.\ 2009, Earth and Planetary Science Letters, 288, 44 

\bibitem[Connelly et al.(2012)]{cbk12} Connelly, J.~N., 
Bizzarro, M., Krot, A.~N., et al.\ 2012, Science, 338, 651

\bibitem[Clayton \& Nittler(2004)]{cn04} Clayton, D.~D., \& Nittler, L.~R.\ 2004, \araa, 42, 39 

\bibitem[Cody et al.(2008)]{cay08} Cody, G.~D., Alexander, C.~M.~O., Yabuta, H., et al.\ 2008, Earth and Planetary Science Letters, 272, 446 

\bibitem[Dohmen et al.(2002)]{dcb02} Dohmen, R., Chakraborty, 
S., \& Becker, H.-W.\ 2002, \grl, 29, 2030

\bibitem[Dohmen \& Chakraborty(2007)]{dc07} Dohmen, R., \& Chakraborty, S.\ 2007, Physics and Chemistry of Minerals, 34, 409 
\bibitem[Floss et al.(2006)]{fsb06} Floss, C., Stadermann, F.~J., Bradley, J.~P., et al.\ 2006, \gca, 70, 2371 
\bibitem[Floss \& Stadermann(2009)]{fs09} Floss, C., \& Stadermann, F.\ 2009, \gca, 73, 2415 
\bibitem[Floss \& Stadermann(2012)]{fs12} Floss, C., \& Stadermann, F.~J.\ 2012, Meteoritics and Planetary Science, 47, 992 
\bibitem[Grossman \& Brearley(2005)]{gb05} Grossman, J.~N., \& Brearley, A.~J.\ 2005, Meteoritics and Planetary Science, 40, 87 

\bibitem[Hirashita \& Yan(2009)]{hy09} Hirashita, H., \& Yan, H.\ 2009, \mnras, 394, 1061 
\bibitem[Hirashita et al.(2010)]{hnyk10} Hirashita, H., Nozawa, T., Yan, H., \& Kozasa, T.\ 2010, \mnras, 404, 1437
\bibitem[Hirashita et al.(2016)]{hna16} Hirashita, H., Nozawa, T., Asano, R.~S., \& Lee, T.\ 2016, arXiv:1602.07094 

\bibitem[H{\"o}fner(2008)]{h08} H{\"o}fner, S.\ 2008, \aap, 491, L1 

\bibitem[Hoppe et al.(2015)]{hlk15} Hoppe, P., Leitner, J., \& Kodol{\'a}nyi, J.\ 2015, ApJL, 808, L9

\bibitem[Huss(1990)]{h90} Huss, G.~R.\ 1990, \nat, 347, 159 
\bibitem[Huss \& Lewis(1995)]{hl95} Huss, G.~R., \& Lewis, R.~S.\ 1995, \gca, 59, 115 
\bibitem[Huss et al.(2006)]{hrg06} Huss, G.~R., Rubin, A.~E., \& Grossman, J.~N.\ 2006, Meteorites and the Early Solar System II, 
eds. D. S. Lauretta and H. Y. McSween Jr. (Tucson, AZ: Univ. Arizona Press), 567 

\bibitem[Hynes \& Gyngard(2009)]{hg09} Hynes, K.~M., \& Gyngard, F.\ 2009, Lunar and Planetary Science Conference, 40, 1198 

\bibitem[Krot et al.(2007)]{kyh07} Krot, A.~N., Yurimoto, H., Hutcheon, I.~D., et al.\ 2007, \gca, 71, 4342 

\bibitem[Krot et al.(2015)]{kna15} Krot, A.~N., Nagashima, K., 
Alexander, C.~M.~O., et al.\ 2015, Asteroids IV, ed. P. Michel et al. (Tucson, AZ: Univ.
Arizona Press), 635  

\bibitem[Leitner et al.(2012)]{lvh12} Leitner, J., Vollmer, C., Hoppe, P., \& Zipfel, J.\ 2012, \apj, 745, 38 
\bibitem[Mathis et al.(1977)]{mrn77} Mathis, J.~S., Rumpl, W., \& Nordsieck, K.~H.\ 1977, \apj, 217, 425 
\bibitem[Messenger et al.(2003)]{mksw03} Messenger, S., Keller, L.~P., Stadermann, F.~J., Walker, R.~M., \& Zinner, E.\ 2003, Science, 300, 105

\bibitem[Miyamoto et al.(1982)]{mft82} Miyamoto, M., Fujii, N.,
 \& Takeda, H.\ 1982, Lunar and Planetary Science Conference Proceedings, 12, 1145 

\bibitem[Miyamoto et al.(2002)]{mma02} Miyamoto, M., Mikouchi, T., \& Arai, T.\ 2002, Antarctic Meteorite Research, 15, 143 
\bibitem[Nguyen et al.(2007)]{nsz07} Nguyen, A.~N., Stadermann, F.~J., Zinner, E., et al.\ 2007, \apj, 656, 1223 
\bibitem[Nguyen et al.(2010)]{nns10} Nguyen, A.~N., Nittler, L.~R., Stadermann, F.~J., Stroud, R.~M., \& Alexander, C.~M.~O.\ 2010, \apj, 719, 166

\bibitem[Nittler et al.(1998)]{naw98} Nittler, L.~R., 
Alexander, C.~M.~O., Wang, J., \& Gao, X.\ 1998, \nat, 393, 222 

\bibitem[Nittler et al.(2013)]{nas13} Nittler, L.~R., Alexander, C.~M.~O., \& Stroud, R.~M.\ 2013, Lunar and Planetary Science Conference, 44, 2367 
\bibitem[Nozawa \& Fukugita(2013)]{nf13} Nozawa, T., \& Fukugita, M.\ 2013, \apj, 770, 27 

\bibitem[Nozawa et al.(2007)]{nkh07} Nozawa, T., Kozasa, T., Habe, A., et al.\ 2007, \apj, 666, 955 
\bibitem[Nozawa et al.(2015)]{nwh15} Nozawa, T., Wakita, S., Hasegawa, Y., \& Kozasa, T.\ 2015, ApJL, 811, L39 

\bibitem[Opeil et al.(2010)]{ocb10} Opeil, C.~P.,
 Consolmagno, G.~J., \& Britt, D.~T.\ 2010, \icarus, 208, 449 

\bibitem[Sears et al.(1980)]{sgmr80} Sears, D.~W., Grossman, J.~N., Melcher, C.~L., Ross, L.~M., \& Mills, A.~A.\ 1980, \nat, 287, 791
\bibitem[Scott \& Jones(1990)]{sj90} Scott, E.~R.~D., \& Jones, R.~H.\ 1990, \gca, 54, 2485 
\bibitem[Scott et al.(1994)]{sjr94} Scott, E.~R.~D., Jones, R.~H., \& Rubin, A.~E.\ 1994, \gca, 58, 1203 

\bibitem[Wakita \& Sekiya(2011)]{ws11} Wakita, S., \& Sekiya, M.\ 2011, Earth, Planets, and Space, 63, 1193 
\bibitem[Wakita et
 al.(2014)]{wni14} Wakita, S., Nakamura, T., Ikeda, T., \& Yurimoto, H.\ 2014, Meteoritics and Planetary Science, 49, 228 

\bibitem[Yasuda \& Kozasa(2012)]{yk12} Yasuda, Y., \& Kozasa, T.\ 2012, \apj, 745, 159 

\bibitem[Yomogida 
\& Matsui(1983)]{ym83} Yomogida, K., \& Matsui, T.\ 1983, Meteoritics, 18, 430

\bibitem[Zinner(2003)]{z03} Zinner, E.~K.\ 2003, Treatise
 on Geochemistry, 1, 711 
\end{thebibliography}
\end{document}